\documentclass{PoS}
\pdfoutput=1
\usepackage{cite}
\usepackage{subfig}
\usepackage{wrapfig}
\usepackage{sidecap}

\newcommand\beq{\begin{eqnarray}}
\newcommand\eeq{\end{eqnarray}}
\newcommand\bal{\begin{align}}
\newcommand\eal{\end{align} }

\newcommand{\mybar}[1]%
        {\kern 0.6pt\overline{\kern -0.6pt#1\kern -0.6pt}\kern 0.6pt}

\title{Lattice study of trapped fermions at unitarity}

\ShortTitle{Lattice study of trapped fermions at unitarity}

\author{\speaker{Amy N. Nicholson}\\
       Institute for Nuclear Theory, University of Washington, Seattle, WA 98195-1550\\
       E-mail: \email{amynn@u.washington.edu}}

\author{Michael G. Endres\\
        Physics Department, Columbia University, New York, NY 10027, USA and\\
        Theoretical Physics Laboratory, RIKEN, Wako, Saitama 351-0198, Japan\\
        E-mail: \email{endres@riken.jp}}

\author{David B. Kaplan\\
        Institute for Nuclear Theory, University of Washington, Seattle, WA 98195-1550\\
        E-mail: \email{dbkaplan@phys.washington.edu}}

\author{Jong-Wan Lee\\
        Institute for Nuclear Theory, University of Washington, Seattle, WA 98195-1550\\
        E-mail: \email{jwlee823@u.washington.edu}}

\abstract{We present a lattice study of up to $N=20$ unitary fermions confined to a harmonic trap. Our preliminary results show better than 1\% agreement with high precision solutions to the many-body Schrodinger equation for up to $N=6$. We are able to make predictions for larger $N$ which were inaccessible by the Hamiltonian approach due to computational limitations. Harmonic traps are used experimentally to study cold atoms tuned to a Feshbach resonance. We show that they also provide certain benefits to numerical studies of many-body correlators on the lattice. In particular, we anticipate that the methods described here could be used for studying nuclear physics.}

\FullConference{The XXVIII International Symposium on Lattice Field Theory\\
                June 14-19,2010\\
                Villasimius, Sardinia Italy}

\begin{document}

\section{Introduction}

The experimental study of trapped, ultracold atoms has provided impetus to explore such systems computationally, particularly near the limit of unitarity. In this regime, all scales relevant to the interaction vanish, corresponding to a zero range two-particle interaction with an infinite s-wave scattering length. This insensitivity to the scale of the interaction allows one to extract physical quantities which are universal in nature. Thus, in addition to being directly relevant to cold atom experiments, numerical studies of this system serve as a stepping stone to more complicated systems such as nuclei, where s-wave scattering lengths are unnaturally large compared to the range of the interaction \cite{Kaplan:1998we, Kaplan:1998tg}.

A recently developed lattice method for describing unitary fermions \cite{Endres:2010gs} allows us to explore several issues of wide interest in lattice field theory: in particular, how to extract the properties of conformal systems from calculations at finite lattice spacing, volume, and particle density, and how to construct optimal interpolating fields for strongly interacting, many-fermion systems. In addition, we demonstrate that it is possible to obtain a very clean signal for up to $N=20$ fermions using modest computational resources.

Our new method is presented in detail in companion proceedings \cite{Endres:2010gs}. In these proceedings, we report a preliminary study of up to $N=20$ unitary fermions trapped in a harmonic potential. We can benchmark our method and systematic errors for up to $N=6$ against high precision solutions to the many-body Schrodinger equation, achieving agreement at the $1\%$ level. We believe this is the first microscopic study to explore higher $N$ without invoking a variational principle or requiring costly importance sampling. We will then present preliminary results for the Bertsch parameter ($\xi$) and the pairing gap. In another companion proceedings, we will present results for untrapped unitary fermions \cite{Lee:2010gs}.

\section{Lattice Theory}

Details of our lattice theory, along with a discussion of numerical costs and benefits, can be found in \cite{Endres:2010gs}. Our approach is a highly improved version of the theory presented in \cite{Chen:2003vy} at zero chemical potential and zero temperature. As outlined in \cite{Endres:2010gs}, we are able to employ a simple iterative procedure for propagator production, whose form is 
\beq
K^{-1}(\tau;0)=D^{-1}X(\tau-1)D^{-1}\cdots D^{-1}X(0)D^{-1} ,
\eeq
where $D$ is the kinetic part of the transfer matrix and $X$ represents the four-fermion interaction, which is produced stochastically through an auxiliary $Z_2$ field. This interaction is highly tuned to produce $p$ cot $\delta_0 \ll 1$, where $\delta_0$ is the s-wave scattering phase shift, for two fermions in a finite box up to arbitrarily high momenta.

We include an external potential through the addition of a field-independent term to our interaction operator:
\beq
X(\tau) \rightarrow X(\tau) + e^{-\frac{\kappa}{2} \sum_{i=1}^3 (x_i-L/2)^2} -1,
\eeq
This introduces $\mathcal{O}(b_{\tau})$ errors, where $b_{\tau}$ is the temporal lattice spacing \footnote{In a future publication we will adopt the form $X(\tau)\rightarrow e^{-\frac{\kappa}{2} \sum_{i=1}^3 (x_i-L/2)^2} X(\tau) e^{-\frac{\kappa}{2} \sum_{i=1}^3 (x_i-L/2)^2}$, which has discretization errors of $\mathcal{O}(b_{\tau}^2)$.}. We will discuss systematic errors in the next section.

To build our sources, we begin with the set of single particle wavefunctions, $|n_i\rangle$, for free fermions in a harmonic trap, where $n_i = \{ (0,0,0), (1,0,0),(0,1,0),\cdots\}$ in the Cartesian basis. While a simple Slater determinant of such states is sufficient to find the ground state for the trapped system, a more sophisticated approach involving two-body correlations gives superior overlap. To build pairing into the system we use the solution for two particles at unitary in a trap, 
\beq
\langle x,y|\psi_{PAIR}\rangle \propto \frac{e^{-(x^2+y^2)/(2L_0^2)}}{|x-y|} , 
\eeq
where $L_0$ is the characteristic trap size, defined in the next section. In order to avoid an $\mathcal{O}(V)$ increase in computation time, we use single particles states at the source and two particle states at the sink to form an $(N/2)\times(N/2)$ Slater determinant as in \cite{Carlson:2003zz}:
\beq
C_N(\tau)=\langle det(S_{ij}) \rangle, \hspace{5mm} S_{ij}=\langle \psi_{PAIR} | K^{-1}(t;0) | n_i n_j \rangle ,
\eeq
where $| n_i n_j \rangle$ represents the direct product $|n_i\rangle \otimes |n_j\rangle$. For odd N, the two-body state in the highest energy level is replaced by the corresponding single particle state. 

\section{Systematic Errors}

The relevant scales of the trapped system inside a periodic box are the box size, $L$, the characteristic size of the trap, $L_0 = (\kappa M)^{-1/4}$, and the oscillator frequency, $\omega = \sqrt{\kappa/M}$, where $\kappa$ and $M$ are the spring constant and fermion mass respectively. To reach the continuum and infinite volume limits we require $b_s \ll L_0 \ll L$, where $b_s$ is the spatial lattice spacing, and $b_{\tau} \ll \omega$. Thus, our systematic errors will be controlled by the dimensionless quantities $\omega b_{\tau}$, $b_s/L_0$, and $L_0/L$. To balance the need for small temporal discretization errors with the computational cost associated with the number of time steps required to reach the ground state, we have chosen $\omega b_{\tau} = 0.0136$ for this preliminary study.

\begin{figure}
\centering
\begin{minipage}[t!]{0.56\textwidth}
\includegraphics[width=\linewidth]{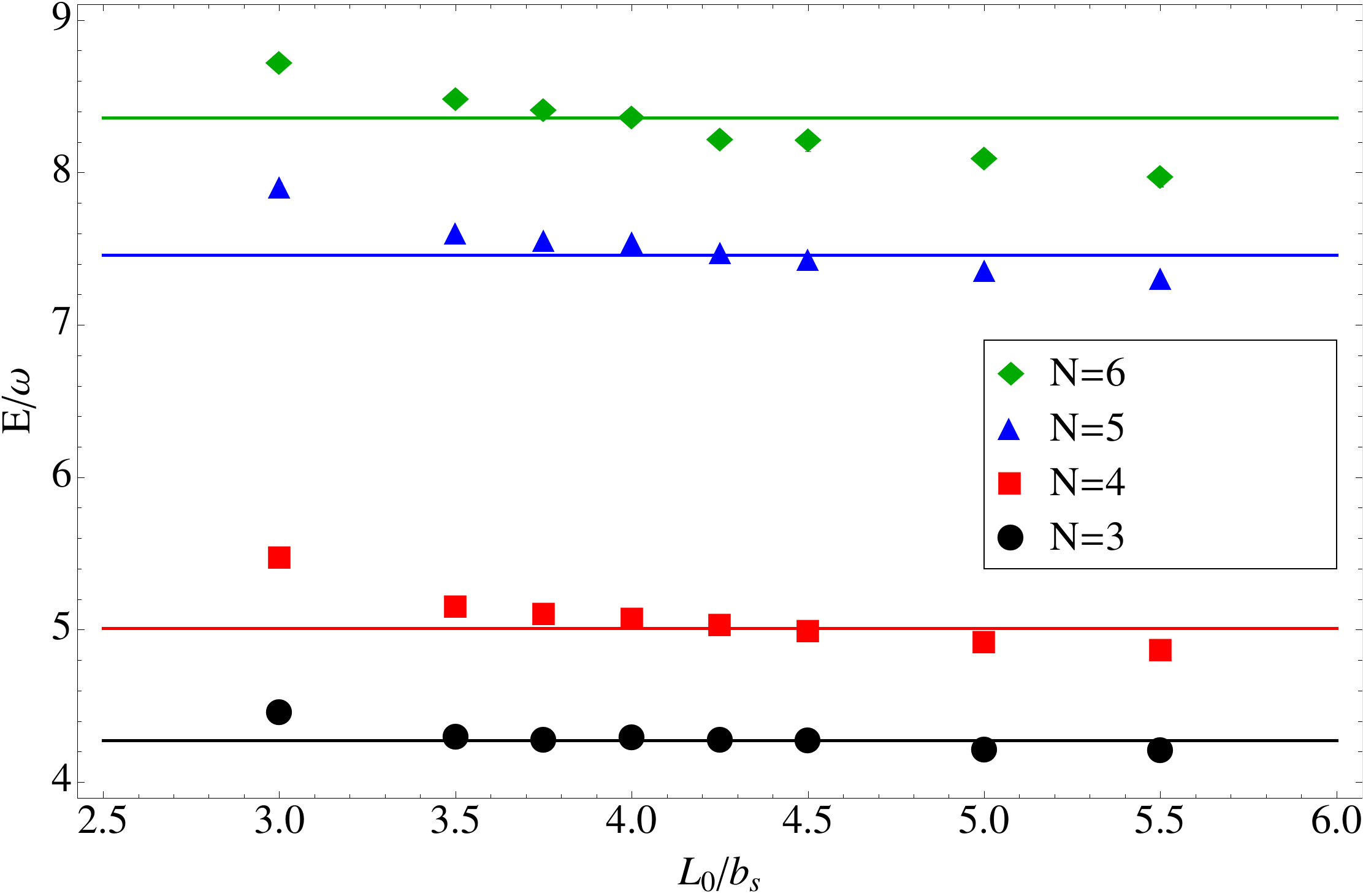}
\caption{\label{fig:SysErrors}Ground state energies as a function of $L_0/b_s$ at fixed $L/b_s=32$ (left) and $L/b_s$ at fixed $L_0/b_s=4$ (right) for various $N$. Solid lines are results from \cite{2010arXiv1008.3191B}.}
\end{minipage}
\hspace{8pt}
\begin{minipage}[t!]{0.4\textwidth}
\centering
\includegraphics[width=6cm]{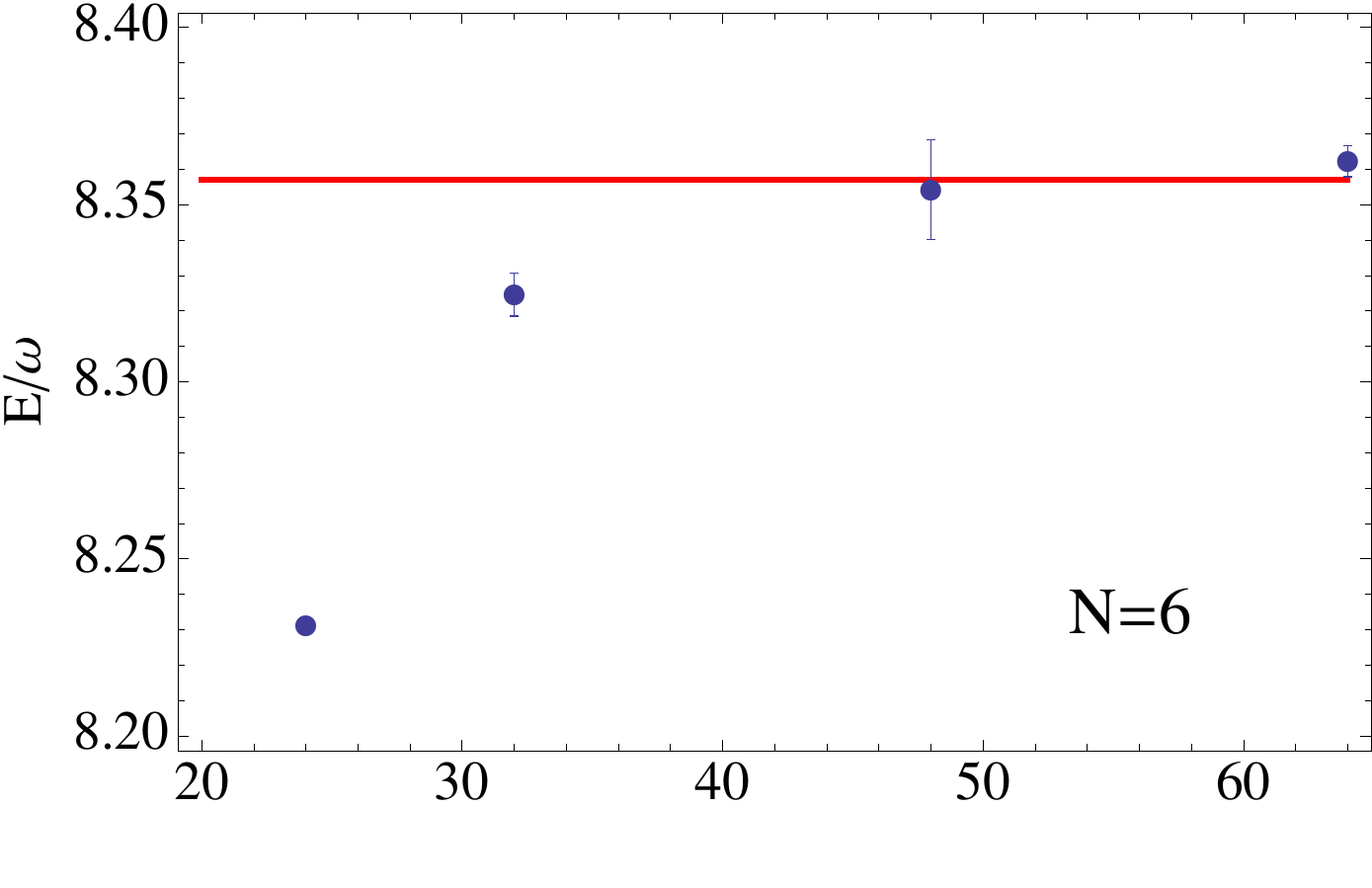}
\includegraphics[width=6cm]{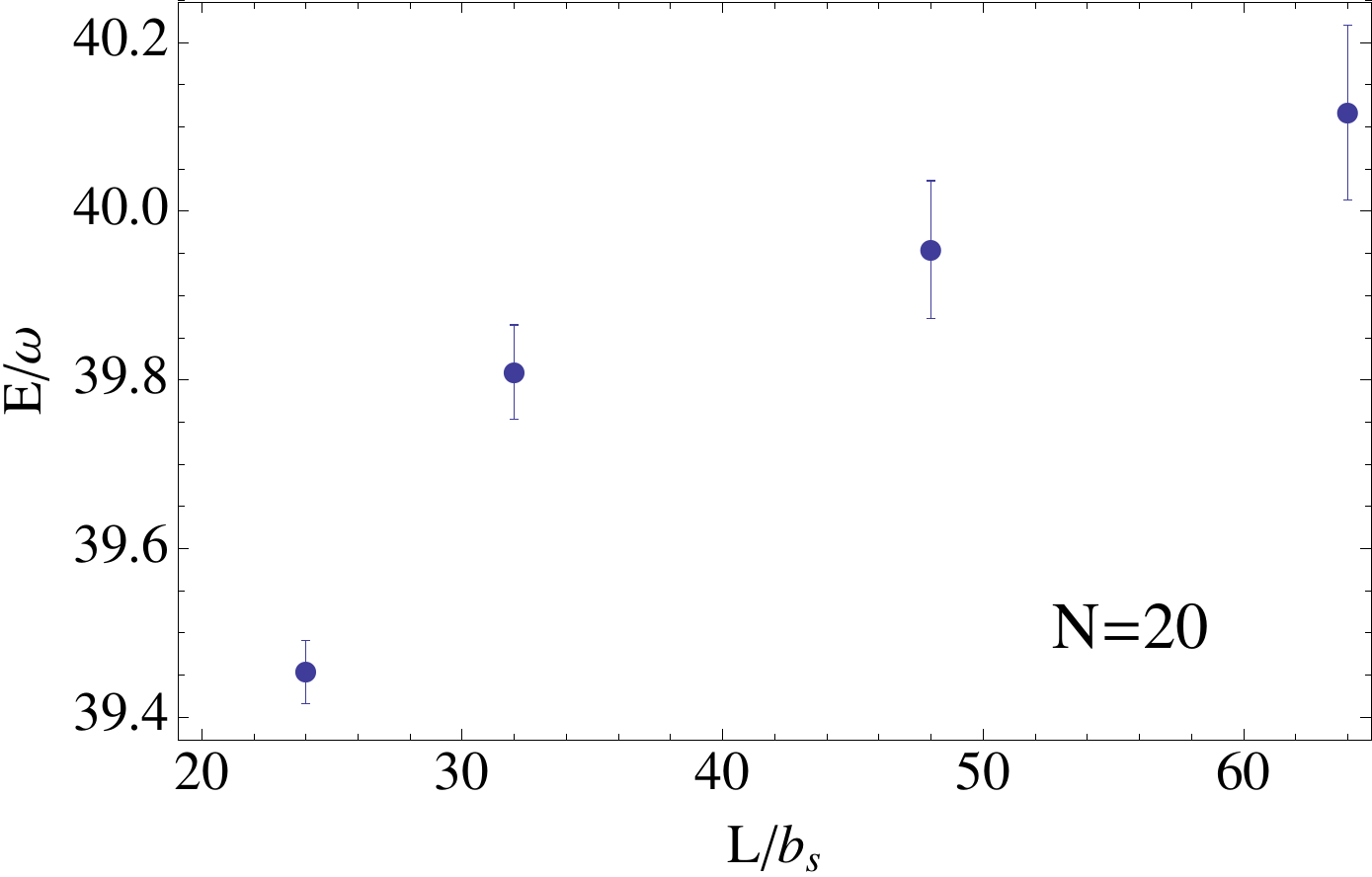}
\end{minipage}
\end{figure}

For a given box size, the choice of $L_0$ must take into account both discretization errors and finite volume errors. The expectation is that for small $L_0/b_s$ spatial discretization errors will dominate. Since the discretization is implemented as a hard cutoff in momentum space, the sensitivity of the state to the infinite potential at the boundary results in an increase in the associated energy. Conversely, for large $L_0/b_s$ finite volume errors will dominate. The periodic boundary conditions in space result in attractive interactions from image particles, causing a decrease in energy. Thus, a scan of the ground state energies as a function of $L_0/b_s$ is necessary to determine at which value we can minimize both types of error. Fig.~\ref{fig:SysErrors} (left) presents our findings.

To ensure small systematic errors for larger N, we have chosen a value of $L_0/b_s=4.0$, and performed the calculation at $L/b_s=24, 32, 48, 64$ (Fig.~\ref{fig:SysErrors}, right). Based on our results for different volumes, we expect that for all $N$ used in this study, at $L/b_s=64$ the finite volume errors should be less than $\sim 0.5\%$.

\section{Results}

\begin{figure}[h]
\begin{center}
\includegraphics[width=6cm]{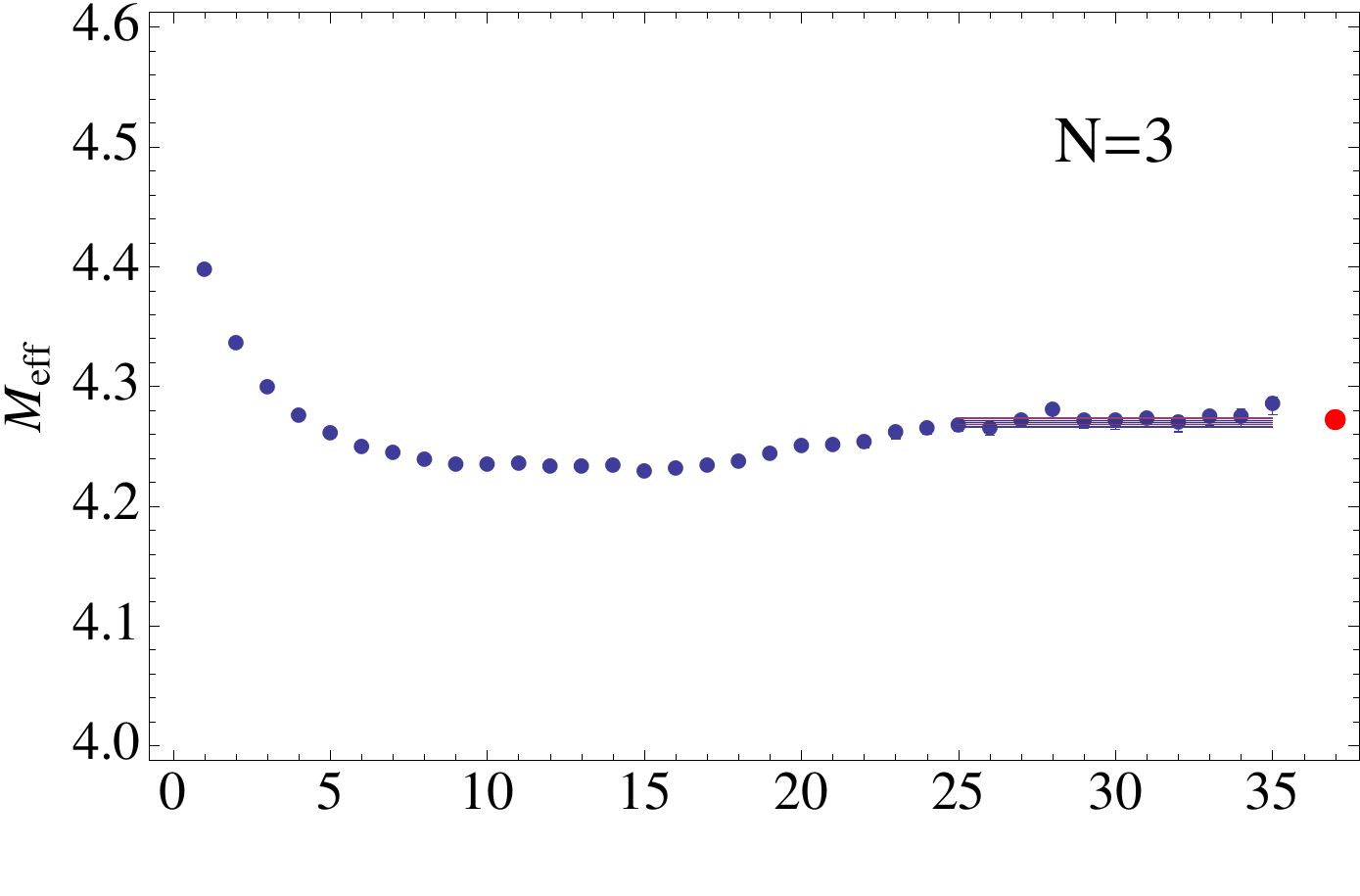}
\includegraphics[width=6cm]{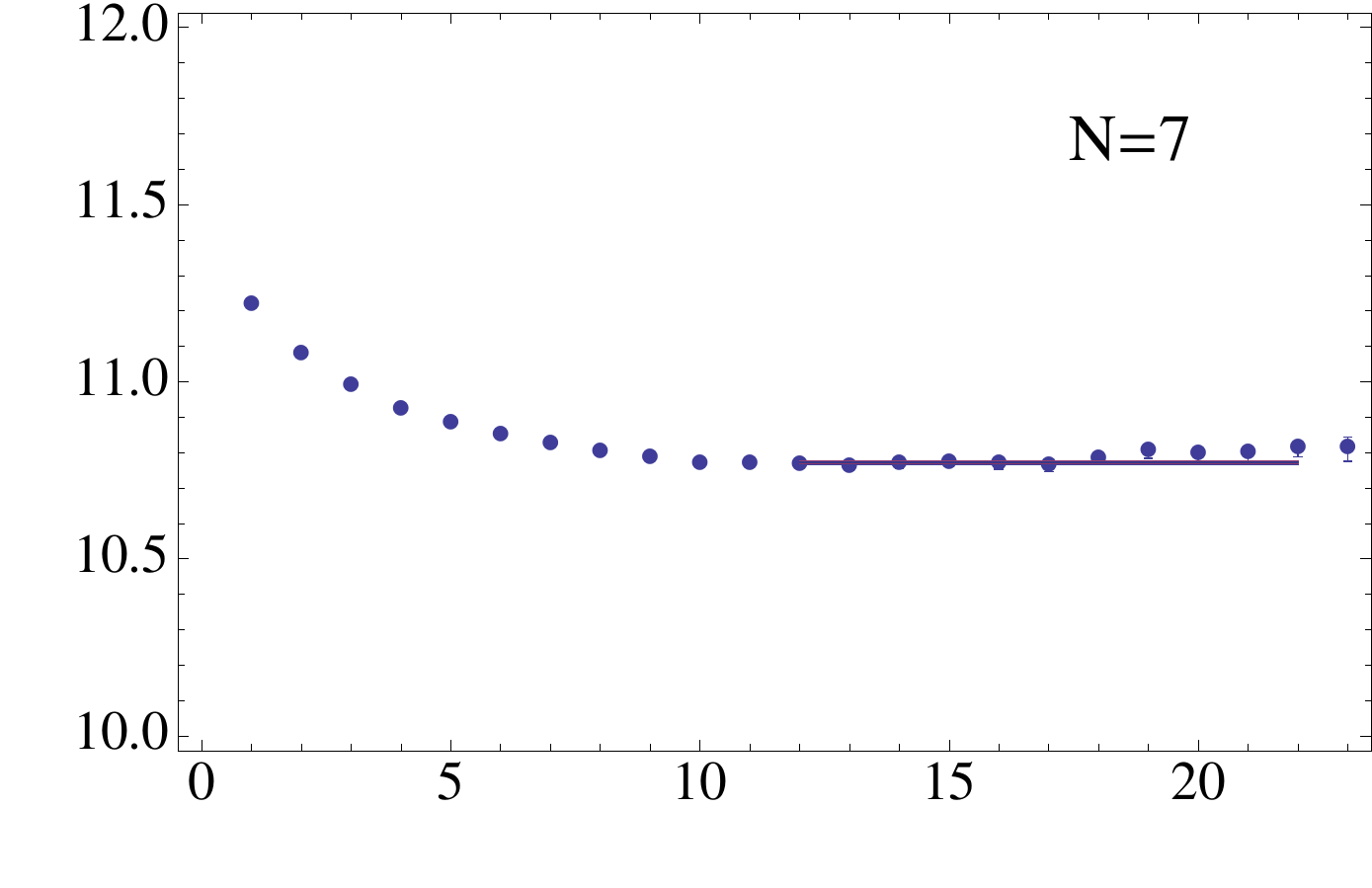}

\includegraphics[width=6cm]{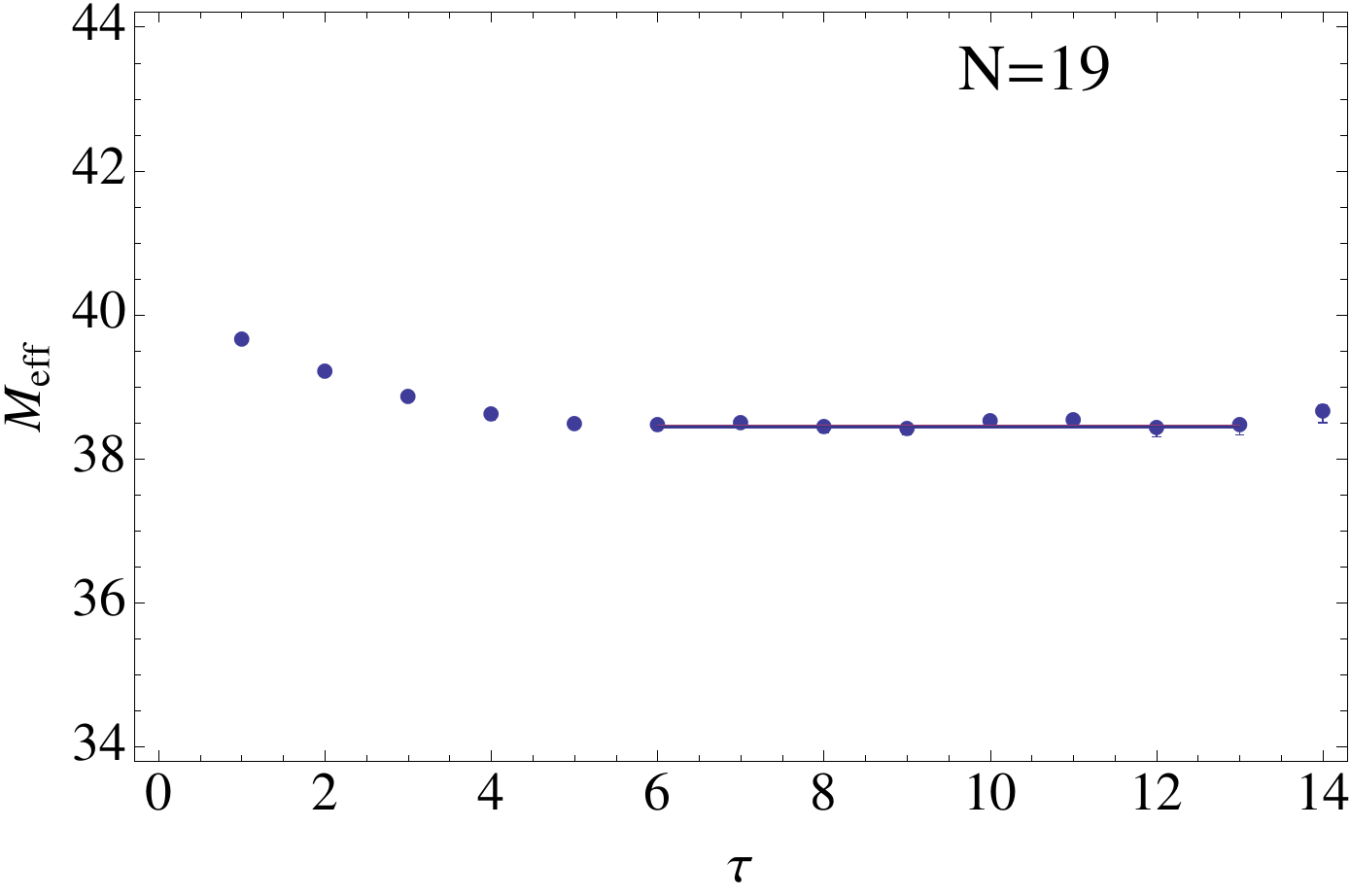}
\includegraphics[width=6cm]{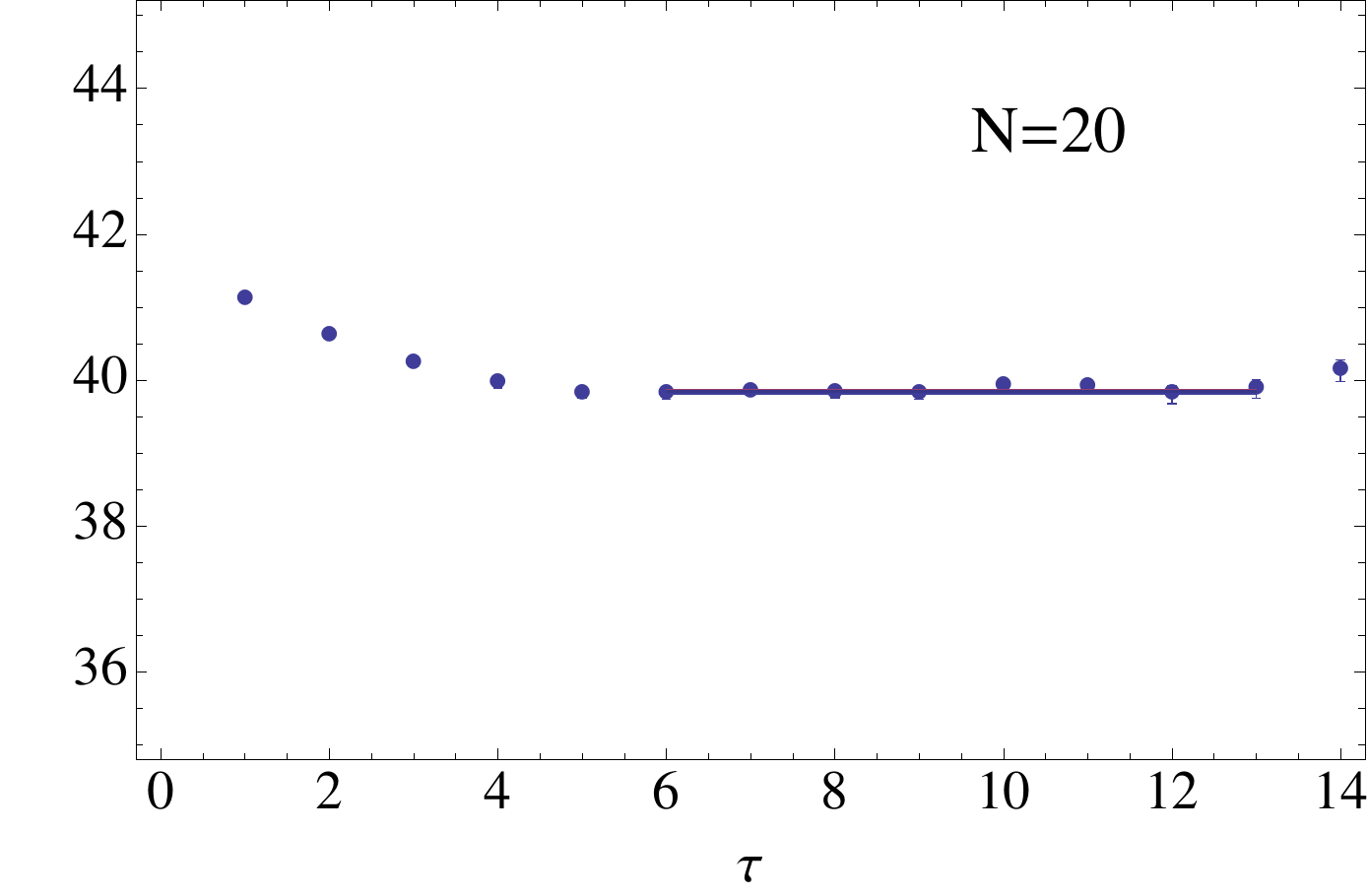}
\caption{\label{fig:effMass}Effective mass plots and fits (solid lines) for various $N$. The red point is the exact ground state energy for $N=3$.}
\end{center}
\end{figure}

For up to $N=20$ fermions, we have generated approximately $10$ million configurations for each of the four volumes, using a total of $350$ thousand CPU hours. We have found excellent signal for all $N$, with little signal-to-noise problem (Fig.~\ref{fig:effMass}). As discussed in \cite{Lee:2010gs}, we find that results for unitary fermions in the absence of a trap are far more difficult to obtain. We believe that the spatial confinement enforced by the harmonic trap provides numerical benefits to lattice calculations of interacting systems, in particular early plateaus and far less sensitivity to the choice of interpolating operator. 

\begin{figure}
\centering
\includegraphics[width=0.7\textwidth]{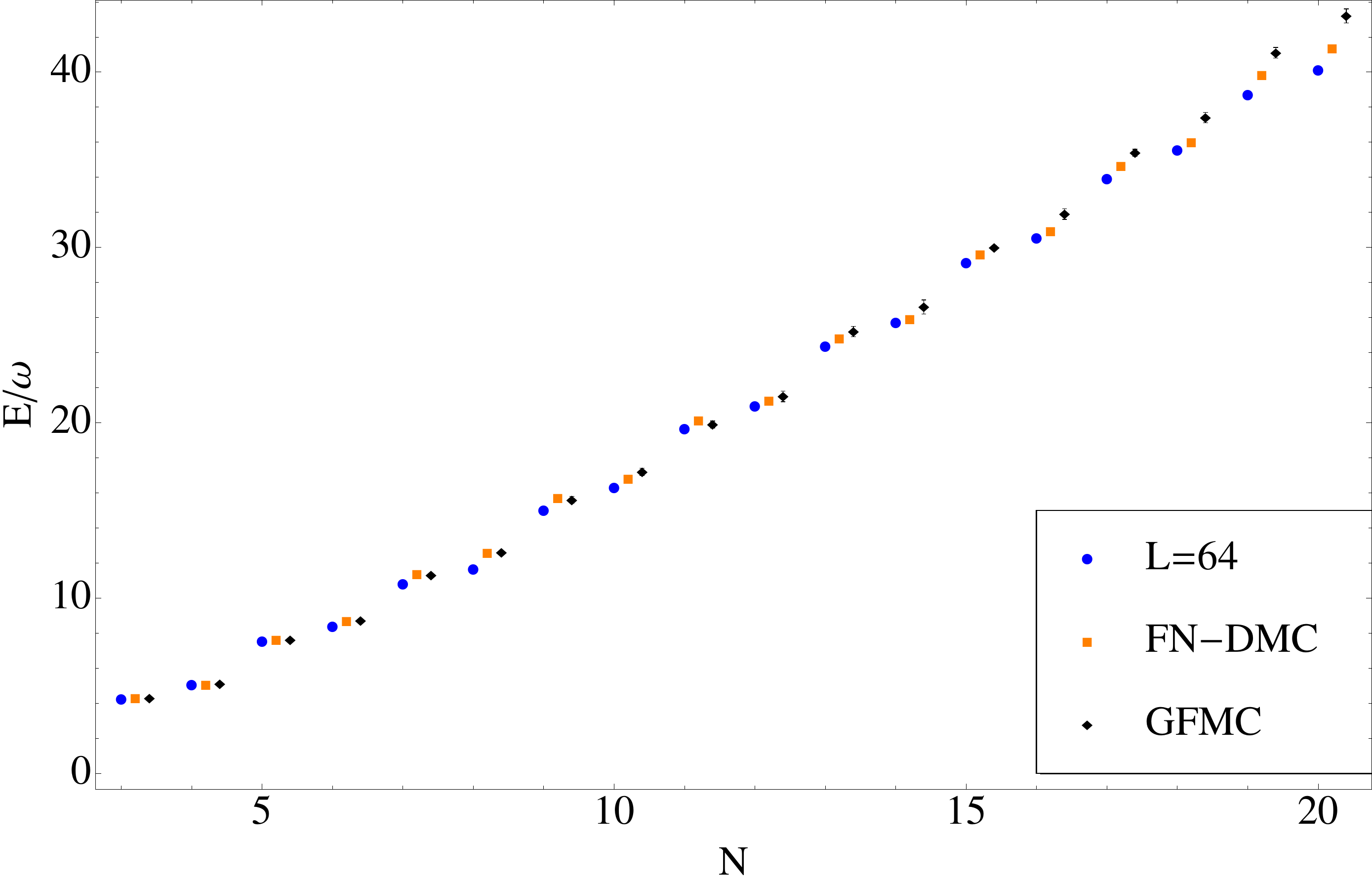}
\caption{Ground state energies as a function of $N$ in units of $\omega$ (preliminary). For comparison, we show results from GFMC \cite{Chang:2007zz} and FN-DMC \cite{2007PhRvL99w3201B}.}
\label{fig:eall}
\end{figure}

To extract the ground state energies, we resampled the data using the bootstrapping technique. We then performed correlated $\chi^2$ fits to the plateau regions. Our fitting systematic errors were found by varying the endpoints of the fitting region by $\Delta \tau=2$. Our results are presented in Fig.~\ref{fig:eall}. We also show the results from two fixed-node calculations: a Green's function Monte Carlo (GFMC) approach \cite{Chang:2007zz} and a Diffusion Monte Carlo (FN-DMC) approach \cite{2007PhRvL99w3201B}. By using the fixed-node constraint along with a variational principle, both of these methods can only provide upper bounds on the ground state energies. We find that our energies are consistently lower than those obtained using both of these methods, and that this discrepancy grows with N, with $N=20$ being approximately 7\% lower than the GFMC approach, and 3\% lower than FN-DMC. Our values and their corresponding statistical and systematic errors are reported in Table~\ref{tab:eall}.

\begin{table}[b!]\small
\begin{center}
\begin{tabular}{|c|c|c|c|c|c|c|c|c|}
\hline
$N$ & $E/\omega$ & $E/\omega$ (FN-DMC) &N & $E/\omega$ &N & $E/\omega$ &N & $E/\omega$ \\
\hline
3 & 4.232(2)(3) & 4.273 & 8 & 11.6491(70)(1) & 13 & 24.35(2)(3)& 18 & 35.540(20)(3) \\
4 & 5.058(1)(1)&5.009 &9 &  14.990(9)(4)& 14 &25.70(2)(9)& 19 &38.69(7)(14) \\
5 &7.513(3)(2) & 7.458&10 &16.289(9)(41)  & 15 &29.11(2)(5) & 20 &40.12(7)(4)\\
6 & 8.362(3)(12)& 8.358& 11& 19.65(1)(4) & 16 &30.52(2)(2) &  & \\
7 & 10.797(7)(5) & & 12& 20.95(2)(5) & 17 & 33.91(3)(4)&  & \\
\hline
 \end{tabular}
 \caption{Ground state energies as a function of $N$ in units of $\omega$ (preliminary). The first error is statistical while the second is a systematic error from the choice of fitting region. For $N\leq6$ we also present high-precision results from \cite{2010arXiv1008.3191B}.}
 \label{tab:eall}
 \end{center}
\end{table}

By performing correlated fits to the data for different $N$ we can extract two additional quantities of interest to very high precision. The first is the quantity known as the Bertsch parameter. For both unitary and non-interacting fermions in a box, the only relevant scale is the density, thus the ratio of their ground state energies must be a constant, $\xi$. With a trap present one can show, using the Local Density Approximation (LDA), that the energy of the interacting system in the thermodynamic limit is given by \cite{2005PhRvA..72d1603P}:
\beq
E_{int}(N,\omega) = \sqrt{\xi} E_{free}(N,\omega) .
\eeq

\begin{figure}
\begin{minipage}[t]{0.486\textwidth}
\centering
\includegraphics[width=\textwidth]{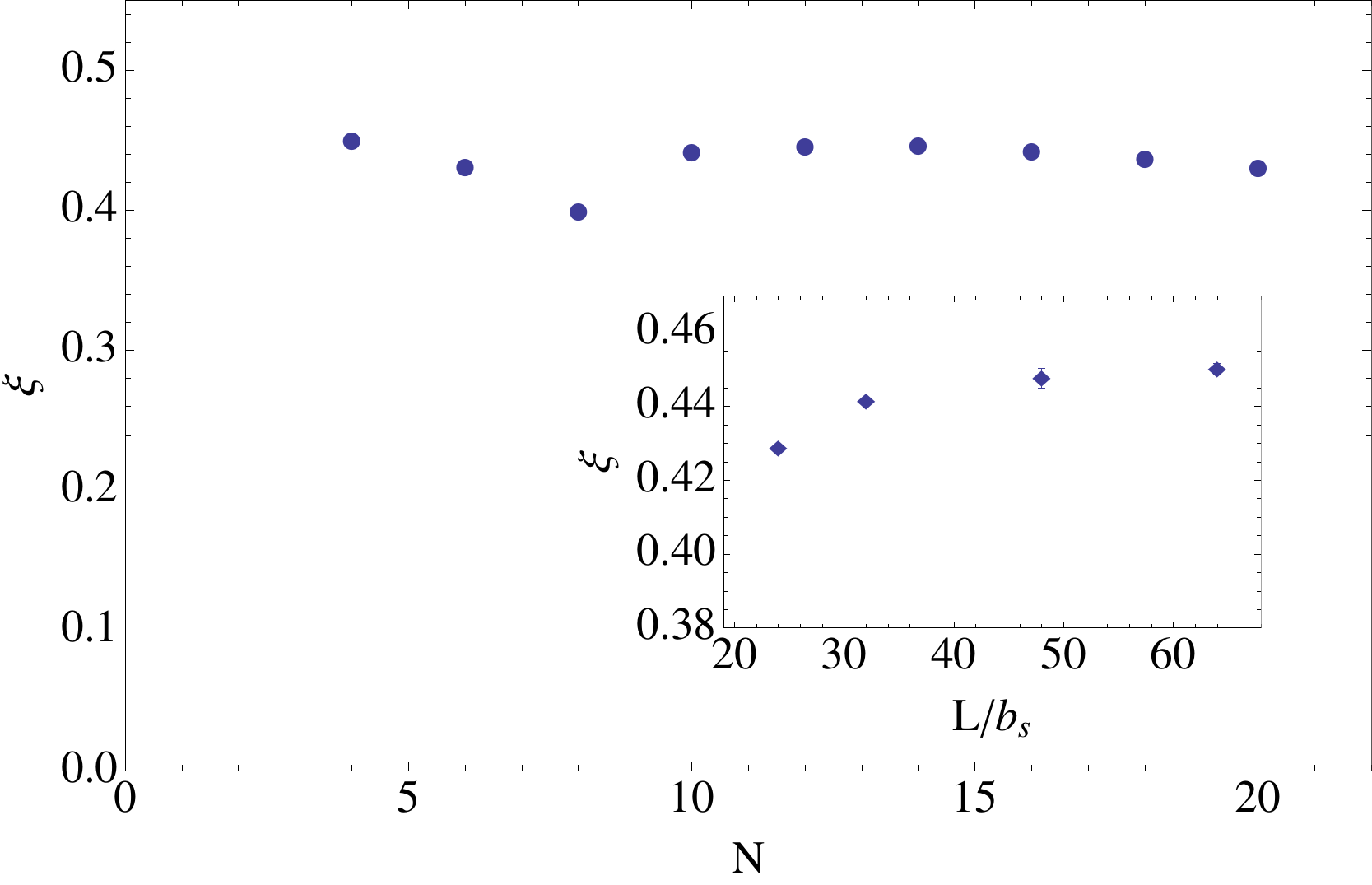}
\caption{$\xi = \left[E_{int}(N,\omega)/E_{free}(N,\omega)\right]^2$ as a function of $N$. Error bars are smaller than data points. Inset: Constant fit to the second shell as a function of $L/b_s$.}
\label{fig:bertsch}
\end{minipage}
\hspace{8pt}
\begin{minipage}[t]{0.486\textwidth}
\centering
\includegraphics[width=\textwidth]{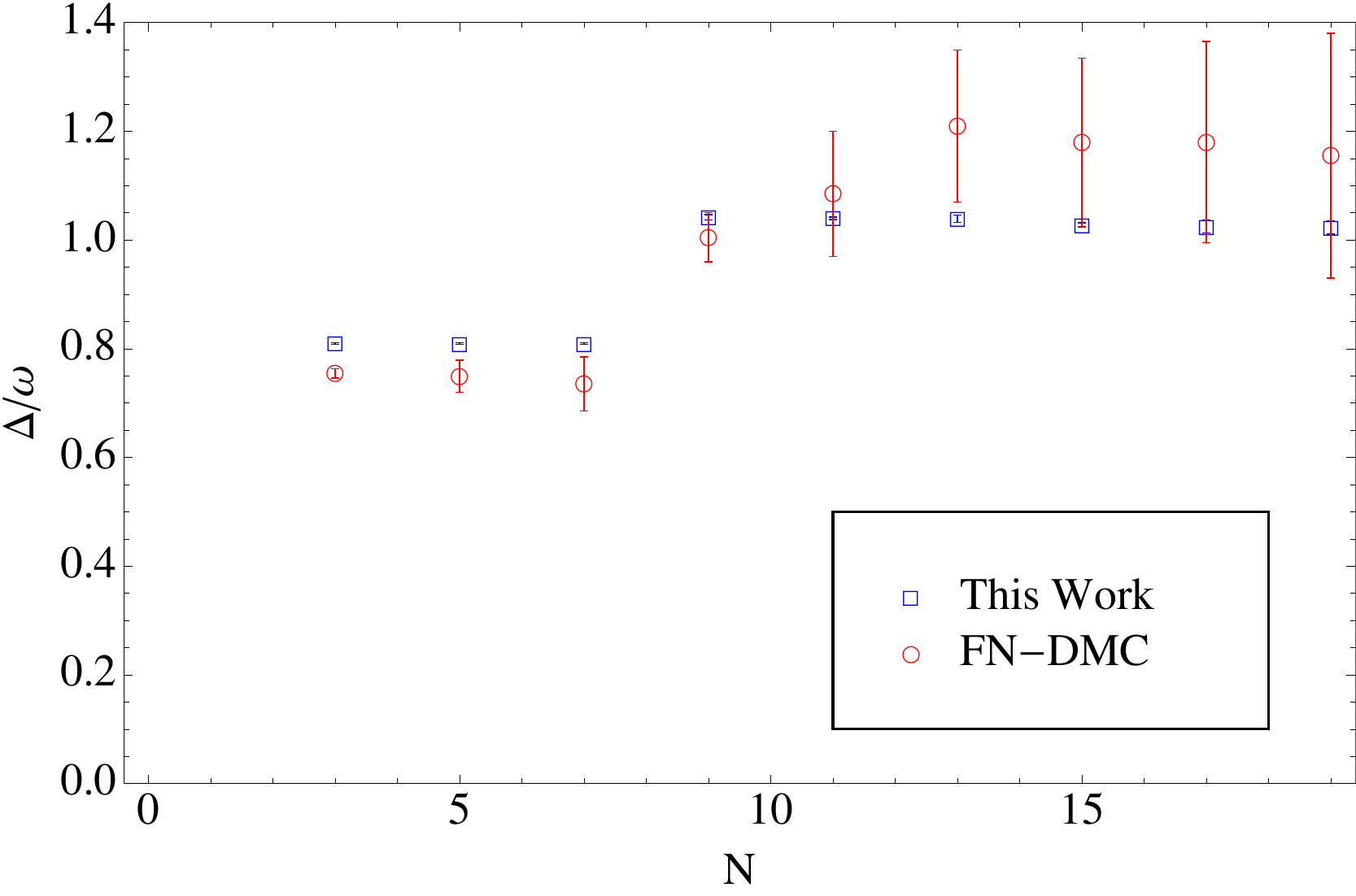}
\caption{The pairing gap as a function of $N$. Error bars are smaller than data points. For comparison, the results from FN-DMC \cite{2007PhRvL99w3201B}.}
\label{fig:gap}
\end{minipage}
\end{figure}

Our results as a function of $N$ are shown in Fig.~\ref{fig:bertsch} for $L/b_s=64$. One sees clear shell structure for the full range of $N$ considered. This indicates that we are probably not sufficiently close to the thermodynamic limit to extract an accurate Bertsch parameter. The inset shows correlated fits to the second shell as a function of $L/b_s$, with our preliminary value from the largest volume being $\xi = 0.450(1)$. For comparison, the data from GFMC and FN-DMC give $\xi = 0.50$ and $\xi=0.46$, respectively. While the Bertsch parameter we extract is lower than that from variational parameter calculations, the value is still higher than calculations with no external potential and extractions from experiment. For example, a recent GFMC calculation finds $\xi=0.40(1)$ \cite{2008PhRvC..77c2801G}, while an experimental study finds $\xi=0.39(2)$ and $\xi=0.41(2)$ using two different extraction methods \cite{2009JLTP1541L}. Similarly, using our lattice method we find a preliminary value of $0.412(4)$ for the untrapped system \cite{Lee:2010gs}. This could indicate a slower convergence to the thermodynamic limit for the trapped case, where the LDA must be invoked. In future studies, we plan to go to higher $N$ to obtain a more accurate prediction for the Bertsch parameter.

The second quantity of interest is the pairing gap, 
\beq
\Delta(N) = E(N) - \frac{1}{2} \left[ E(N-1)-E(N+1)\right]  \quad \textrm{ (for N odd)}.
\eeq
The gap quantifies half of the energy required to break a fermion pair. We present our results as a function of $N$ in Fig.~\ref{fig:gap}. For comparison, we also show the results from \cite{2007PhRvL99w3201B}. By performing a correlated fit in $N$ we are able to achieve statistical errors which are an order of magnitude smaller than the previous calculation. The gap shows a clear shell structure for the range of $N$ considered. Due to this shell structure, we will need to extend our calculation to much larger $N$ before we are able verify the $N^{1/9}$ behavior proposed in \cite{2007arXiv0707.1851S}.

\section{Conclusions}

We have performed a study of unitary fermions in a trap which match high precision results to within 1\%. We have shown that the signal-to-noise problem for our lattice method is well under control, and predict that results for at least $N=40$ will be numerically feasible. In addition, we have performed a preliminary study of systematic errors.

In a future publication, in addition to extending the study to larger $N$, we will improve the discretization of our potential, introduce a Galilean invariant interaction to ensure that $p$ cot $\delta_0 = 0$ for boosted pairs of particles, and perform an even more extensive analysis of systematic errors. Our ability to precisely tune $p$ cot $\delta_0$ suggests future applications to nuclear systems. The use of harmonic traps for extracting nuclear physics has also gained interest recently \cite{Luu:2010hw}. Due to the numerical benefits of including a harmonic trap, our formulation lends itself to such studies.

\acknowledgments{This work was supported by U. S. Department of Energy grants DE-FG02-92ER40699 (to M. G. E.) and DE-FG02-00ER41132 (to D. B. K., J-W. L. and A. N. N.). This research utilized resources at the New York Center for Computational Sciences at Stony Brook University/Brookhaven National Laboratory which is supported by the U.S. Department of Energy under Contract No. DE- AC02-98CH10886 and by the State of New York. The authors would like to thank D. Blume, et al. for granting permission to present their results before publication.}

\bibliographystyle{h-physrev.bst}
 \bibliography{PQQCDv8}

\end{document}